\documentclass[preprint,floats,aps,epsfig,nofootinbib,amssymb]{revtex4}
\usepackage{graphicx}
\usepackage{epsfig}
\usepackage{subfigure}
\usepackage{dcolumn}
\usepackage{bm}
\def\gsim{\lower0.5ex\hbox{$\:\buildrel >\over\sim\:$}}
\def\lsim{\lower0.5ex\hbox{$\:\buildrel <\over\sim\:$}}

\def \Dslash {D\!\!\!\!/}

\begin{document}


\title{The Muon Magnetic Moment in the TeV Scale Seesaw Models}

\author{\bf Wei Chao}
\email{chaowei@ihep.ac.cn}

\affiliation{ Institute of High Energy Physics, Chinese Academy of
Sciences, Beijing 100049, China
\vspace{2.5cm} }

\begin{abstract}
The reported discrepancy of the muon abnormal magnetic moment
$a_\mu^{}$ has impacts on the low energy phenomenology. In this
paper we calculate the corrections to $a_\mu^{}$ in the standard
model extended by the TeV scale seesaw models. We show that the
correction induced by the type-I seesaw model is negative and of the
order ${\cal O} (10^{-11})$, which can be neglected compared with
$a_\mu^{\rm SM}$. The correction induced by the type-II seesaw
model, which depends on the mass of the Higgs triplet $m_\Delta^{}$
and the Yukawa coupling $Y_\Delta^{}$, can be of the order ${\cal
O}( 10^{-10})$ and compensate for the discrepancy between
$a_\mu^{\rm SM}$ and $a_{\mu}^{\rm exp}$. The correction induced by
the type-III seesaw model is also negative and can be of the order
${\cal O}(10^{-10})$.
\end{abstract}
\maketitle

\section{Introduction}
For a spin $1/2$ particle, the relation between its magnetic moment
and its spin reads $\vec{\mu}=g(e/2m)\vec{s}$. The Dirac equation
predicts for the gyromagnetic factor $g=2$, but radiative
corrections to the lepton-photon-lepton vertex in quantum field
theory may switch the value slightly. The abnormal magnetic moment
is then defined as $a=(g-2)/2$.

There has been a long history in measuring and calculating the muon
abnormal magnetic moment $a_\mu^{}$. In particular the steadily
improving precision of both the measurements and the predictions of
$a_\mu$ and the disagreement observed between the two have made the
study of $a_\mu$ one of the most active research fields in particle
physics in recent years. The final result of the ``Muon g-2
Experiment"(E821) for $a_\mu^{}$ reads \cite{g-2}
\begin{eqnarray}
a_\mu^{\rm exp}=(11659208\pm6)\times10^{-10} \ ,
\end{eqnarray}
which deviates from the standard model (SM) prediction:
\begin{eqnarray}
\Delta a_\mu^{}=a_\mu^{\rm exp}-a_\mu^{\rm SM}=22(10) \times
10^{-10}\ .
\end{eqnarray}
Many new physics scenarios have been proposed to interpret the
non-vanishing and positive value of $\Delta a_\mu^{}$\cite{mg-22}.
Meanwhile new physics proposed to solve some other problems may
potentially contribute to $\Delta a_\mu^{}$.

On the neutrino sector, the discovery of neutrino oscillations has
confirmed the theoretical expectation that neutrinos are massive and
lepton flavors are mixed, providing the first evidence for physics
beyond the SM in particle physics. The most appealing and natural
idea for generating small neutrino masses is the seesaw mechanisms
\cite{Seesaw1, typeII, typeiii}, which rely on the existence of
heavy particles such as right-handed Majorana neutrinos, triplet
scalar or triplet fermions. A salient feature of the seesaw
mechanisms is that the thermal leptogenesis mechanism \cite{FY} can
work well to account for the cosmological baryon number asymmetry. A
direct test of the seesaw mechanisms would involve the detection of
those heavy particles at a collider and the measurement of their
Yukawa couplings with the electroweak doublets. If such Yukawa
couplings are similar to the other fermion Yukawa couplings, the
masses of those heavy particles turn out to be too high to be
experimentally accessible.

To submit to the experiment, some kinds of TeV scale seesaw models
\cite{Pilaftsis, early, smirnov, Han, tev type-II} were proposed, in
which the masses of the heavy particles are set at the electroweak
scale. The key point of such seesaw scenarios is to adjust the
structures of heavy particles' Yukawa couplings to guarantee that
$M_\nu^{}$ (i.e., the mass matrix of light Majorana neutrinos)
equals to zero at the tree level. Then tiny but non-vanishing
neutrino masses can be ascribed to slight perturbations or radiative
corrections to $M_\nu^{}$ in the next-to-leading order
approximation. A prominent feature of such kinds of seesaw scenarios
is that the interactions of heavy particles with the SM gauge
bosons\footnote{In the type-I seesaw mechanism, $N$ can interact
with the SM gauge bosons and Higgs through its mixing with the light
SM SU(2) $\nu_{\rm L}^{}$.} and Higgs are not necessary suppressed,
leading to very interesting lepton-number-violating phenomenology
mediated by heavy particles at high-energy colliders such as the
Tevatron, the LHC and the ILC.

TeV scale seesaw scenarios may lead to large unitarity violation of
the lepton mixing matrix (MNS). However, a global analysis of
current neutrino oscillation data and precision electroweak data
yields very stringent constraints on the non-unitarity of the MNS
matrix. Therefore a systematic investigation of the low energy
phenomenology induced by such seesaw scenarios is necessary and
important.

In this paper, we will calculate the corrections to $a_\mu^{}$
induced by the TeV scale heavy particles. We show that the
correction to $a_\mu^{}$ induced by heavy Majorana neutrinos is of
the order $-{\cal O}(10^{-11})$ in the type-I seesaw model.
Corrections induced by the doubly charged Higgs boson and singly
charged Higgs boson can be of the order ${\cal O}(10^{-10})$ in the
type-II seesaw model. Therefore $\Delta a_\mu^{}$ can be completely
saturated by $\Delta a_\mu^{\rm II}$. Whereas, the correction
induced by triplet fermions is of the order $-{\cal O}(10^{-10})$ in
the type-III seesaw model.

The outline of the paper is as follows. In section II we describe
some basics of the TeV scale seesaw scenarios. Section III is
devoted to the calculation of corrections to $a_\mu^{}$ induced by
various seesaw models. Some conclusions are drawn in section IV.

\section{Some basics of the TeV scale seesaw models}
We regularize our notations and conventions in this section by
reviewing some basics of the TeV seesaw scenario. After gauge
symmetry spontaneous breaking, the  neutrino mass terms turn out to
be
\begin{eqnarray}
-{\cal L}_{\rm mass} = \frac{1}{2} \overline{\left( \nu^{}_{\rm L}
~N^c_{\rm R}\right)} \left( \matrix{ M^{}_{\rm L} & M^{}_{\rm D} \cr
M^T_{\rm D} & M^{}_{\rm R}}\right) \left( \matrix{ \nu^c_{\rm L} \cr
N^{}_{\rm R}}\right) + {\rm h.c.} \; ,
\end{eqnarray}
where $\nu^c_{\rm L} \equiv C \overline{\nu^{}_{\rm L}}^T$ with $C$
being the charge conjugation matrix, likewise for $N^c_{\rm R}$. The
overall $6\times 6$ neutrino mass matrix in ${\cal L}^{}_{\rm
mass}$, denoted as ${\cal M}$, can be diagonalized by the unitary
transformation ${\cal U}^\dagger {\cal M} {\cal U}^* = \widehat{\cal
M}$; or explicitly,
\begin{eqnarray}
\left(\matrix{V & R \cr S & U}\right)^\dagger \left( \matrix{
M^{}_{\rm L} & M^{}_{\rm D} \cr M^T_{\rm D} & M^{}_{\rm R}}\right)
\left(\matrix{V & R \cr S & U}\right)^*  = \left( \matrix{
\widehat{M}^{}_\nu & {\bf 0} \cr {\bf 0} & \widehat{M}^{}_{\rm
N}}\right) \; ,
\end{eqnarray}
where $\widehat{M}^{}_\nu = {\rm Diag}\{m^{}_1, m^{}_2, m^{}_3\}$
and $\widehat{M}^{}_{\rm N} = {\rm Diag}\{M^{}_1, M^{}_2, M^{}_3\}$
with $m^{}_i$ and $M^{}_i$ (for $i=1, 2, 3$) being the light and
heavy Majorana neutrino masses, respectively. Note that the $3\times
3$ rotation matrices $V$, $U$, $R$ and $S$ are non-unitary, but they
are correlated with one another due to the unitarity of $\cal U$.

In the basis where the flavor eigenstates of three charged leptons
are identified with their mass eigenstates, the standard
charged-current interactions between $\nu^{}_\alpha$ and $l_L^{}$
(for $l = e, \mu, \tau$) can be written as
\begin{eqnarray}
-{\cal L}^{}_{\rm cc} = \frac{g}{\sqrt{2}} \left[
\overline{l^{}_{\rm L}} V \gamma^\mu \nu^{}_{ i} W^-_{\mu} +
\overline{l^{}_{\rm L}} R \gamma^\mu N^{}_{ i} W^-_\mu \right] +
{\rm h.c.} \; .
\end{eqnarray}
It becomes clear that $V$ describes the charged-current interactions
of three light Majorana neutrinos, while $R$ is relevant to the
charged-current interactions of three heavy Majorana neutrinos. One
can similarly write out the interactions between the Majorana
neutrinos and the neutral gauge boson (or Higgs) in the chosen
flavor basis \cite{Pilaftsis}:
\begin{eqnarray}
{\cal L}_{\rm Z} &=& -\frac{g}{2 c_{\rm W}}  \overline{\nu_{\rm
L}^{}} R
\gamma^\mu P_L^{} N_{i}^{} Z_\mu^{} + {\rm h.c.}  ~, \\
{\cal L}_{\rm H} &=& -\frac{g}{2} \frac{M_{i}^{}}{M_{\rm W}^{}}
\overline{\nu_{\rm L}}^{} R  P_R^{}  N_i^{} h^0  +{\rm  h.c.}  ~.
\end{eqnarray}

There are constraints on the non-unitarity of $VV^\dagger$ from
electroweak decays. Ratios of $\mu$, $\tau$, $W$ and $\pi$ decays,
used often in order to test universality, can be interpreted as
tests of lepton mixing unitarity. They result in constraints for the
diagonal elements of $VV^\dagger$. The lepton-flavor-violating
processes, which occur at the one-loop level, constrain the
off-diagonal elements of $VV^\dagger$. A global fit to the
constraints listed above results in \cite{unitary}
\begin{eqnarray}
VV^\dagger\approx\left(\matrix{0.994\pm 0.005&<7.0\cdot 10^{-3}
&<1.6\cdot10^{-2}\cr <7.0\cdot 10^{-5}&0.995\pm
0.005&<1.0\cdot10^{-2}\cr <1.6\cdot 10^{-2}&<1.0\cdot
10^{-2}&0.995\pm0.005}\right)\ ,
\end{eqnarray}
at the $90\%$ confidence level. It is clear that the deviation of
$VV^\dagger$ from the identity matrix can be as large as a few
percents. Therefore the low energy phenomenology induced by heavy
neutrinos is not negligible.

\section{$a_\mu^{}$ in various seesaw models}
The most general form for the photon-muon vertex function
$\Gamma^\mu$, which is consistent with Lorentz covariance, can be
written as \cite{mg-22, form factor}
\begin{eqnarray}
\bar{u}(p_2^{})\Gamma^\mu
u(p_1^{})&=&\bar{u}(p_2^{})\left[F_1^{}(q^2)\gamma^\mu-{i\over
2m_{\mu}^{}} F_2^{}(q^2)\sigma^{\mu\nu}q_\nu^{}+{1\over m_\mu^{}}
F_3^{}(q^2)q^\mu + \right.\nonumber\\&&\left.\gamma_5^{}(G_1^{}
(q^2)\gamma^\mu-{i\over
2m_\mu^{}}G_2^{}(q^2)\sigma^{\mu\nu}q_\nu^{}+{1\over
m_\mu^{}}G_3^{}(q^2)q^\mu)\right] u(p_1^{})\ ,
\end{eqnarray}
where $q=p_2^{}-p_1^{}$ and $m_\mu^{}$ is the mass of muon. The
anomalous magnetic moment of the muon is related to $\Gamma^\mu$ as
follows: $a_\mu^{}= F_2^{}(0)$.

The SM prediction of $a_\mu^{}$ is generally divided into three
parts: $a_\mu^{\rm SM}=a_\mu^{\rm QED}+a_\mu^{\rm EW} + a_\mu^{\rm
Had}$. The QED part includes all photonic and leptonic ($e, \mu,
\tau$) loops starting with the classic $\alpha/2\pi$ Schwinger
contribution. Loop contributions involving $W^\pm,~ Z$ or Higgs
particles are collectively labeled as $a_\mu^{\rm EW}$. The hadronic
part includes the contributions from the quark and gluon loops.
There are contributions induced by various heavy particle loops,
which are contained in TeV scale seesaw models. We will calculate
them in the following.

\subsection{$a_\mu^{}$  in the type-I seesaw scenario}
Assuming that light but non-zero neutrino masses are generated by
the type-I seesaw mechanism, we need to extend the SM with  three
right-handed Majorana neutrinos. The relevant Lagrangian can be
written as
\begin{eqnarray}
{\cal L}^{}_{\rm I} &=&{\cal L}_{\rm SM}^{}-\overline{l^{}_{\rm L}}
Y^{}_\nu \tilde{H} N^{}_{\rm R} - \frac{1}{2} \overline{N^{c}_{\rm
R}} M^{}_{\rm R} N^{}_{\rm R}  + {\rm h.c.} \; ,
\end{eqnarray}
where $M_{\rm R}^{}$ is masses of the right-handed neutrinos.
Integrating out right-handed Majorana neutrinos results in a light
neutrino Majorana mass matrix of the form: $M_\nu^{}=-v^2 Y_\nu^{}
M_{\rm R}^{-1}Y_\nu^T$. In this model the MNS matrix is non-unitary
and the heavy Majorana neutrinos interact with charged leptons
through their mixing with light neutrinos, which was already shown
in Eq. (5). As a result, the muon abnormal magnetic moment receives
contribution from the heavy Majorana neutrino and $W$ boson loop.
The relevant diagram is shown in Fig. 1 (a), which gives the
following correction to $a_\mu^{}$:

\begin{eqnarray}
\Delta a_\mu^{\rm I}={G_{\rm F}^{}m_{\rm \mu}^2\over 8\sqrt{2}
\pi^2}\left(R R^\dagger\right)_{\mu\mu}^{}\left[I\left(M_{\rm W}^2,
M_{ i}^{2}\right)-{10\over 3 }\right]\ ,
\end{eqnarray}
where $I( M_{\rm W}^{2}, M_{ i}^{2})$  can be written as
\begin{eqnarray}
I(M_{\rm W}^2, M_{ i}^2)=M_{\rm W}^2\int dx {4x^2(x+1)\over m_\mu^2
x^2+x( M_{\rm W}^2-M_{ i}^2-m_\mu^2)+M_{ i}^2}\ .
\end{eqnarray}
Suppose that masses of the right-handed Majorana neutrinos are
degenerate. We plot $\Delta{a}_\mu^{\rm I}$ in Fig. 2 by assumming
that $RR^\dagger\sim 1\%$ and the masses of heavy neutrinos lie in
the range $200 {\rm GeV}\leq M_{1}^{} \leq ~ 500 {\rm GeV}$, which
are potentially accessible at the LHC. We can find from the figure
that $\Delta{a}_\mu^{\rm I}$ is negative and not sensitive to
$M_1^{}$. Besides, $\Delta a_\mu^{\rm I}$ is too small to change the
SM prediction significantly.

\subsection{$\Delta a_\mu^{}$ in the type-II seesaw scenario }

We now proceed to the TeV scale type-II seesaw scenario \cite{tev
type-II}. In this scenario an extra scalar triplet ($Y=2$) together
with some heavy Majorana neutrinos is added to the SM. The most
general Lagrangian for this model is
\begin{eqnarray}
{\cal L}_{\rm II}={\rm Tr}\left[(D^\mu \Delta)^\dagger D_\mu^{}
\Delta\right]-m_\Delta^2{\rm Tr}\left(\Delta^\dagger\Delta\right)-
\frac{1}{2} \overline{l^{}_{\rm L}} Y^{}_\Delta \Delta i\sigma^{}_2
l^c_{\rm L} -\overline{l^{}_{\rm L}} Y^{}_\nu \tilde{H} N^{}_{\rm R}
- \frac{1}{2} \overline{N^{c}_{\rm R}} M^{}_{\rm R} N^{}_{\rm R}  +
{\rm h.c.} \
\end{eqnarray}
where $\Delta$ represents the Higgs triplet. After integrating out
heavy fields and the spontaneous gauge symmetry breaking, one
obtains the effective mass matrix for three light neutrinos:
$M_\nu^{}\approx v_\Delta^{} Y_\Delta^{}-v^2 Y_\nu^{} M_{\rm R}^{-1}
Y_\nu^T $, with $v$ and $v_\Delta^{}$ being the vacuum expectation
values (vev's) of the neutral components of $H$ and $\Delta$,
respectively. The smallness of $M_\nu^{}$ is ascribed to a
significant but incomplete cancellation between $v_\Delta^{}
Y_\Delta^{}$ and $v^2 Y_\nu^{} M_{\rm R}^{-1} Y_\nu^T$ terms. There
are totally seven physical Higgs bosons in this model:
doubly-charged $ \Delta^{++}$ and $\Delta^{--}$, singly-charged
$\delta^+$ and $\delta^-$, neutral $A^0$ (CP-odd), and neutral $h^0$
and $H^0$ (CP-even), where $h^0$ is the SM-like Higgs boson. Doubly
charged Higgs boson and charged lepton loops together with singly
charged Higgs boson and neutrino ($W$ boson) loops may contribute to
$a_\mu^{}$. The relevant diagrams are shown in Fig. 1 (b) $-$(f).
Direct calculation results in
\begin{eqnarray}
a _\mu^\Delta&=&{1\over 16\pi^2}\sum_{\alpha=e, \mu, \tau}^{}
|(Y_\Delta^{})_{\mu \alpha}|^2\left[I_1^{} (m_\alpha^2, m_\Delta^2
)+I_2^{}
(m_\alpha^2, m_\Delta^2)\right]\ ,\\
a^\delta_\mu&=&{1\over 16\pi^2}\sum_\alpha^{} |(Y_\Delta^{})_{\mu
\alpha}^{}|^2 (VV^\dagger)_{\alpha\alpha}^{} I_3^{}( m_\alpha^2,
m_\delta^2)\ ,
\end{eqnarray}
with
\begin{eqnarray}
I_1^{}(m_\alpha^2, m_\Delta^2)&=& m_\mu^2\int dx {x(1-x^2)\over x
m_\Delta^2+(1-x) m_\alpha^2+(x^2-x)m_\mu^2}\ ,\nonumber\\
I_2^{}(m_\alpha^2, m_\Delta^2)&=& m_\mu^2 \int dx
{2x(1-x)^2\over(1-x)
m_\Delta^2+x m_\alpha^2+(x^2-x) m_\mu^2}\nonumber\ ,\\
I_3^{}(m_\alpha^2, m_\delta^2)&=& m_\mu^2\int dx {x(1-x)^2\over(1-x)
m_\delta^2+(x^2-x) m_\mu^2 }\ ,
\end{eqnarray}
where $m_\alpha^{}$ $(\alpha=e, \mu, \tau)$ reads as the mass of the
charged lepton. When writing down Eq. (15), we have ignored the
contributions of diagrams (e) and (f) in Fig. 1, which are
suppressed by the masses of light Majorana neutrinos.The total
corrections to $a_\mu^{}$ motivated by the type-II seesaw scenario
is defined by the sum of Eqs. (11), (14) and (15):
$\Delta{a}_{\mu}^{\rm II}= {a}_\mu^{\Delta}+{a}_\mu^{\delta}+\Delta
a_\mu^{\rm I}$.

Notice that $M_{\rm L}^{}$ can be reconstructed via $M^{}_{\rm L} =
V \widehat{M}^{}_\nu V^T + R \widehat{M}^{}_{\rm N} R^T \approx R
\widehat{M}^{}_{\rm N} R^T$ \cite{tev type-II}, which must be a good
approximation. The element of the Yukawa coupling matrix
$(Y_\Delta^{})$ turns out to be
\begin{eqnarray}
(Y_\Delta^{})^{}_{\alpha \beta} = \frac{\left(M^{}_{\rm
L}\right)^{}_{\alpha \beta}}{v^{}_\Delta} \approx \sum^3_{i=1}
\frac{R^{}_{\alpha i} R^{}_{\beta i} M^{}_i}{v^{}_\Delta} \; ,
\end{eqnarray}
where the subscripts $\alpha$ and $\beta$ run over $e$, $\mu$ and
$\tau$. This result implies that the muon magnetic moment depends on
both $R$ and $M^{}_i$.  $v^{}_\Delta$ may affect the gauge boson
masses in such a way that $\rho \equiv M^2_W/(M^2_Z \cos^2
\theta^{}_{\rm W}) = (v^2 + 2v^2_\Delta)/(v^2 + 4v^2_\Delta)$ holds.
By using experimental constraint on the $\rho$-parameter \cite{PDG},
one gets $\kappa \equiv \sqrt{2} ~v^{}_\Delta /v < 0.01$ and
$v^{}_\Delta < 2.5~{\rm GeV}$. We work in the minimal type-II seesaw
scenario \cite{minimalII} and set $v_\Delta^{}=1 {\rm GeV}$ in our
numerical analysis. Let us parametrize the $3\times 1$ complex
matrix $R$ in terms of three rotation angles and three phase angles
\cite{Xing}: $R = (\hat{s}^*_{14}, c^{}_{14} \hat{s}^*_{24},
c^{}_{14} c^{}_{24} \hat{s}^*_{34})^T$, where $c^{}_{ij} \equiv \cos
\theta^{}_{ij}$ and $\hat{s}^{}_{ij} \equiv e^{i\delta^{}_{ij}}
s^{}_{ij}$ with $s^{}_{ij} \equiv \sin \theta^{}_{ij}$ (for $ij =
14, 24, 34$). Combining all electroweak precision constraints, we
may choose a self-consistent parameter space of three mixing angles:
$s^{}_{14} \approx 0$, $s^{}_{24} \in [0, 0.1]$ and $s^{}_{34} \in
[0, 0.1]$. In Fig. 3 we plot $\Delta{a}^{\rm II}_\mu$ as a function
of $m_\Delta^{}$, setting $R$ to its largest allowed values. The
solid, dotted and dashed lines correspond to $M_1^{}=50, 200, 500~
{\rm GeV}$, separately. The short dotted line corresponds to $\Delta
a_\mu^{}$. It is clear that $\Delta{a}^{\rm II}_\mu$ is proportional
to $M_1^{}$ and the deviation of $a_\mu^{}$ from the SM prediction
may be fully saturated by $\Delta{a}^{\rm II}_\mu$. Suppose that
$m_\Delta^{}$ lies in the range $200~ {\rm GeV}\leq m_\Delta^{}\leq
500~ {\rm GeV}$. The experimental result of $a_\mu^{}$ constrains
the mass of the heavy Majorana neutrino to lie below $310.5$ ${\rm
GeV}$.

\subsection{$a_\mu^{}$ in the type-III seesaw scenario}

Let us calculate $a_ \mu^{}$ in the type-III seesaw scenario
\cite{typeiii}, which extends the SM with $SU(2)_{\rm L}^{}$ triplet
of fermions with zero hypercharge. In this model at least two such
triplets (or one triplet plus one singlet) are necessary in order to
genereate non-vanishing light neutrino masses. The relevant
Lagrangian can be written as
\begin{eqnarray}
{\cal L}_{\rm III}={\rm Tr} [\overline{ \Psi}  i\Dslash
\Psi]-{1\over 2} {\rm Tr}[\overline{ \Psi} m_{\Psi}^{} \Psi^C ]-
\sqrt{2} \overline{\ell_L^{}} \tilde{\phi} Y_{\Psi}^{}\Psi +{\rm
h.c.}\ ,
\end{eqnarray}
where $m_ \Psi^{}$ is the mass of triplet fermion and $\Psi$ can be
written as
\begin{eqnarray}
\Psi=\left(\matrix{\Psi^0_{}/ \sqrt{2}&\Psi^+ \cr \Psi^- &
-\Psi^0/\sqrt{2}}\right)\ .
\end{eqnarray}
Integrating out triplet fermions at the tree level results in a
dimension five effective operator which leads to a light neutrino
Majorana mass matrix of the form: $M_\nu^{\rm III}=-{v^2\over 2}
Y_\Psi^{} m_\Psi^{-1} Y_\Psi^T$. The possibility of testing type-III
seesaw at the LHC is discussed in many articles \cite{iii lhc}, in
which lepton-number-violating and (or) lepton-flavor-violating
signals induced by the triplet fermions are discussed. Singly
charged heavy field $\Psi^-$ may contribute to $a_\mu^{}$. The
relevant diagram is shown in Fig. 1 (g). Direct calculation results
in
\begin{eqnarray}
\Delta  a^{\rm III}_\mu= {1\over 8\pi^2}(Y_{\Psi}^{}
Y_\Psi^{\dagger})_{\mu\mu}^{}I_4^{} (m_{\rm H}^{2}, m_\Psi^2)\ ,
\end{eqnarray}
where
\begin{eqnarray}
I_4^{} (m_{\rm H}^{2}, m_\Psi^2)=m_\mu^2\int dx
{{x(1-x)^2}\over(x-x^2) m_\mu^2+(x-1) m_\Psi^2- x m_{\rm H}^2}\ .
\end{eqnarray}
It is clear that $I_4^{} <0$, which means $\Delta{a}^{\rm
III}_\mu<0$. Suppose that there is structure cancelation in
$M_\nu^{\rm III}$, just like what happens in TeV scale type-I and
type-II seesaw models. Then $Y_\Psi^{} \thicksim 1$ and
$\Delta{a}^{\rm III}_\mu \thicksim -{\cal O}( 10^{-10})$, which is
not negligible but theoretically unfavorable.

In summary, we have evaluated the corrections to $a_\mu^{}$ induced
by heavy Majorana neutrinos, triplet scalar and triplet fermions. To
definitely illustrate the effect of different seesaw scenarios, we
summarize our results in table I.
\begin{table}[htbp]
\centering
\begin{tabular}{|c|l|r|}
\hline TeV scale seesaw models & extra heavy particles &~ ~$\Delta a_\mu^{}$~\\
\hline Type-I seesaw & right-handed  neutrinos & $-{\cal O
}(10^{-11})$ \\ \hline Type-II seesaw & right-handed neutrinos+Higgs
triplet & ${\cal O}(10^{-10})$ \\ \hline Type-III seesaw & triplet
fermions & $-{\cal O}(10^{-10})$ \\ \hline
\end{tabular}
\caption{ The corrections to $a_\mu^{}$ induced by  TeV scale heavy
particles, which are contained in various seesaw models. }
\end{table}

\section{Conclusion}
Motivated by the conjecture that new physics at the TeV scale is
responsible for the electroweak symmetry breaking and origin of
neutrino masses, a series of TeV seesaw models were proposed. These
TeV scale seesaw scenarios, in which  sufficient lepton number
(flavor) violation signals are induced, are testable at the LHC and
(or) ILC. Meanwhile extra heavy particles in these models may induce
interesting low energy phenomena. In this article, we have evaluated
the corrections to $a_\mu^{}$ induced by heavy Majorana neutrinos,
triplet scalar and triplet fermions, which are separately included
in type-I, II and III seesaw models. Our results show that the
correction induced by the heavy neutrinos is ignorable compared with
$a_\mu^{\rm SM}$. Corrections induced by the doubly charged Higgs
boson and singly charged Higgs boson may be of the order ${\cal
O}(10^{-10})$ and $\Delta a_\mu^{}$ can be completely saturated by
$\Delta a_\mu^{\rm II}$. Whereas the correction induced by triplet
fermions can be of the order $-{\cal O} (10^{-10})$, which is
theoretically unfavorable. In conclusion, TeV scale type-II and
type-III seesaw scenarios can significantly contribute to $
a_\mu^{}$. The running of the LHC may potentially verify which
mechanism is responsible for $\Delta a_\mu^{}$ and the origin of
neutrino masses.
\begin{acknowledgments}
The author is indebted to Prof. Z.Z. Xing for polishing up the
manuscript with many suggestions and corrections. This work was
supported in part by the National Natural Science Foundation of
China.
\end{acknowledgments}

\newpage

\begin{figure}[t]
\subfigure[]{\epsfig{file=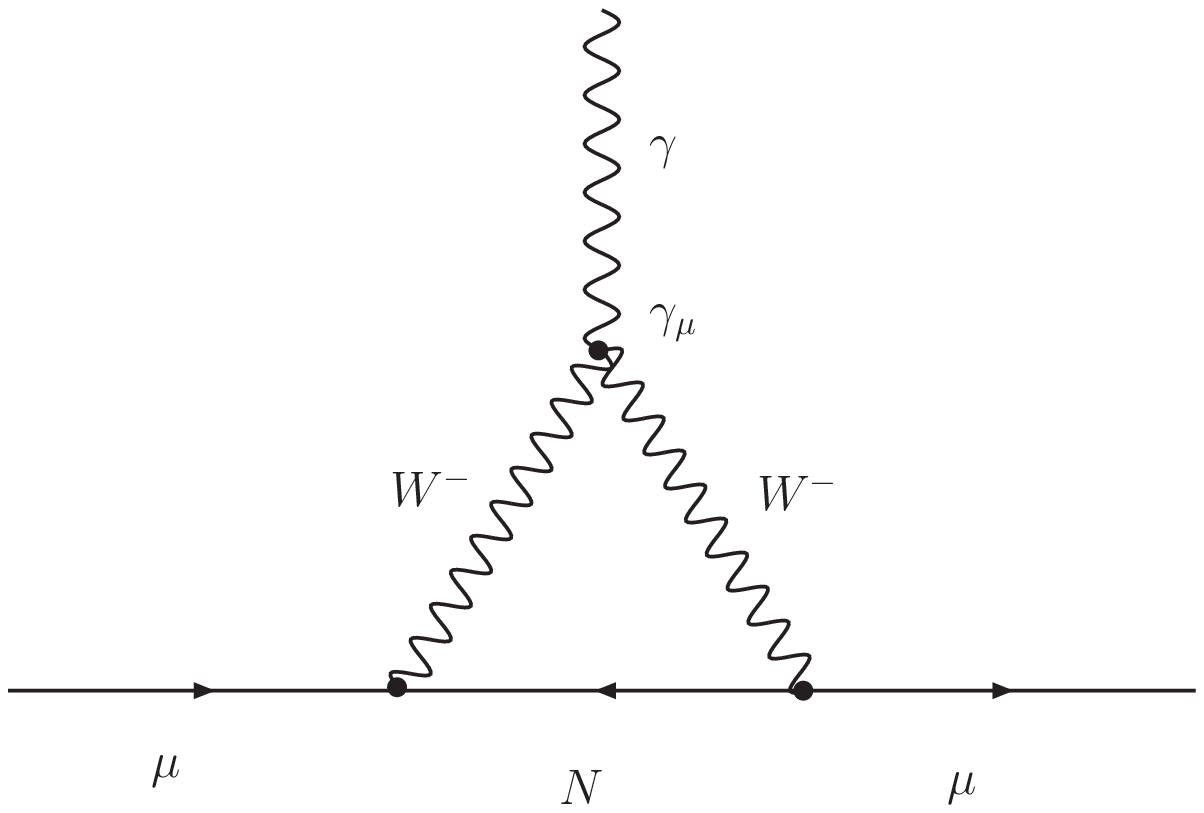,height=4cm,width=5.1cm,angle=0}}
\vspace{0.6cm}
\subfigure[]{\epsfig{file=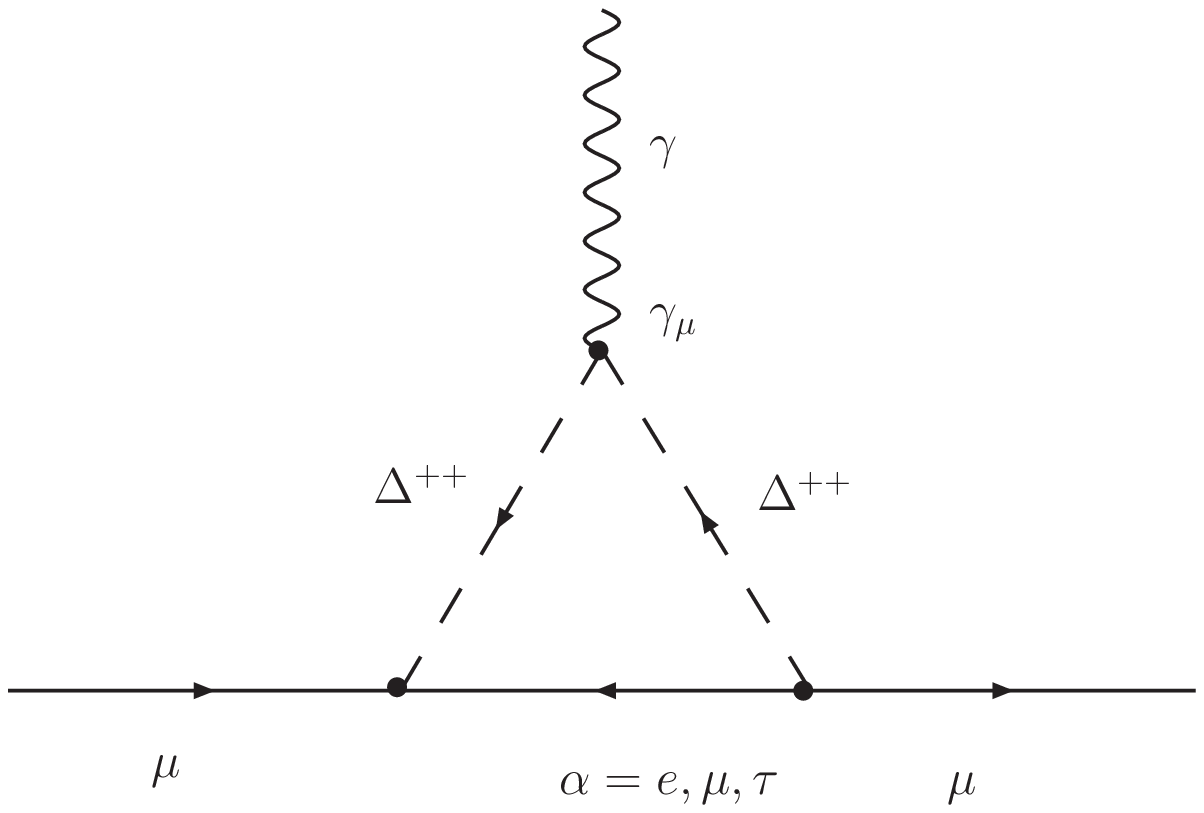,height=4cm,width=5.1cm,angle=0}}
\subfigure[]{\epsfig{file=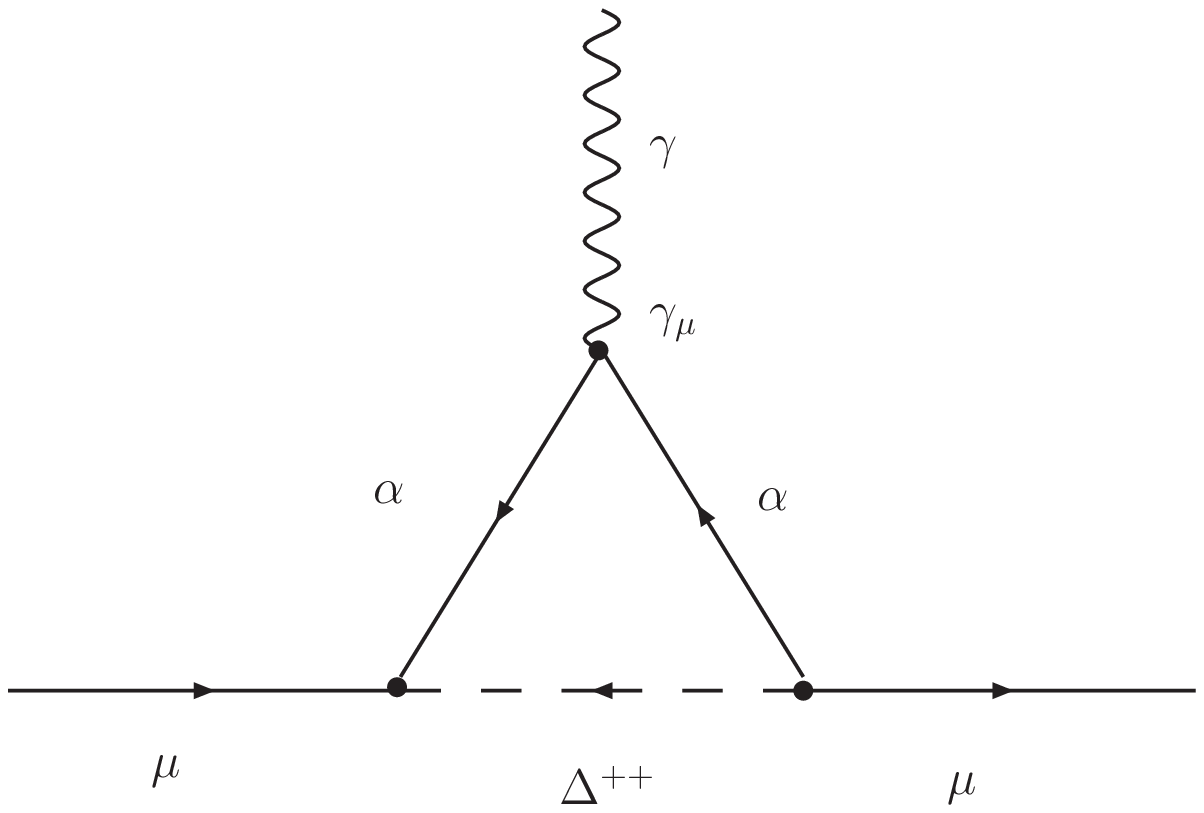,height=4cm,width=5.1cm,angle=0}}
\vspace{0.6cm}
\subfigure[]{\epsfig{file=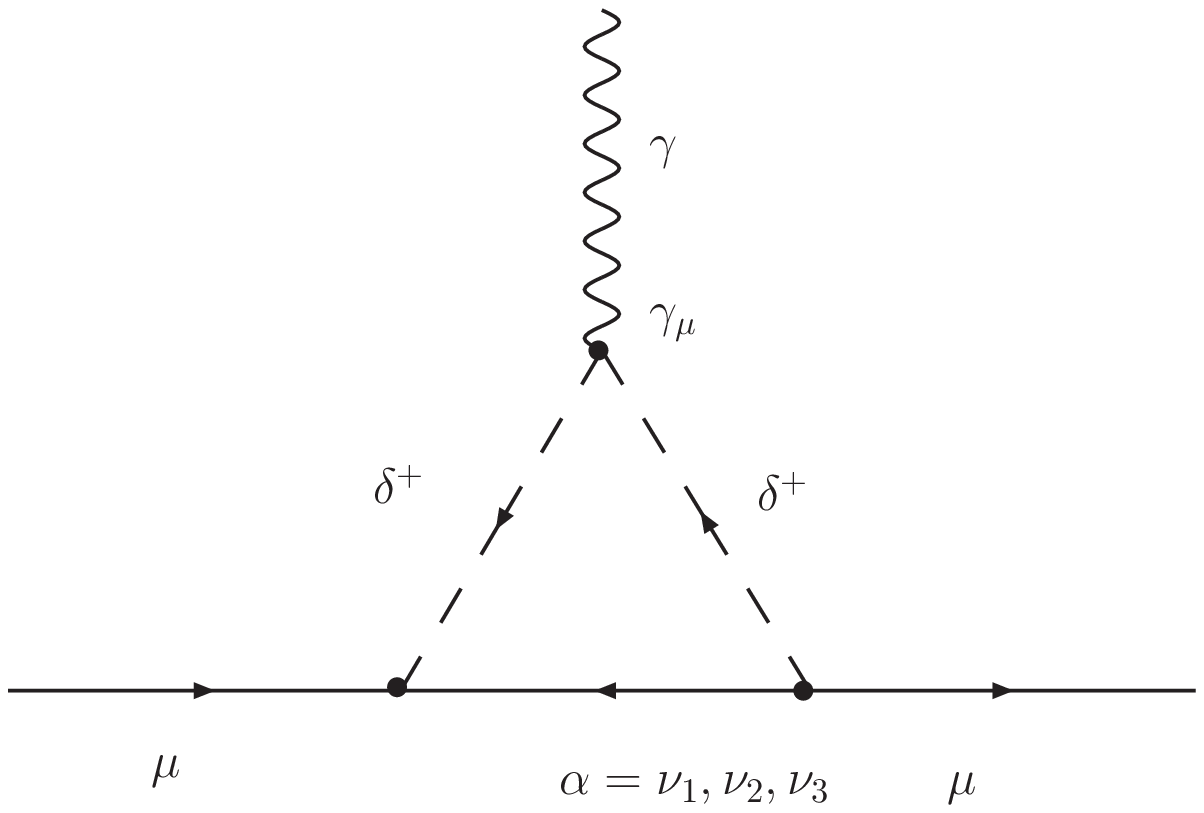,height=4cm,width=5.1cm,angle=0}}
\subfigure[]{\epsfig{file=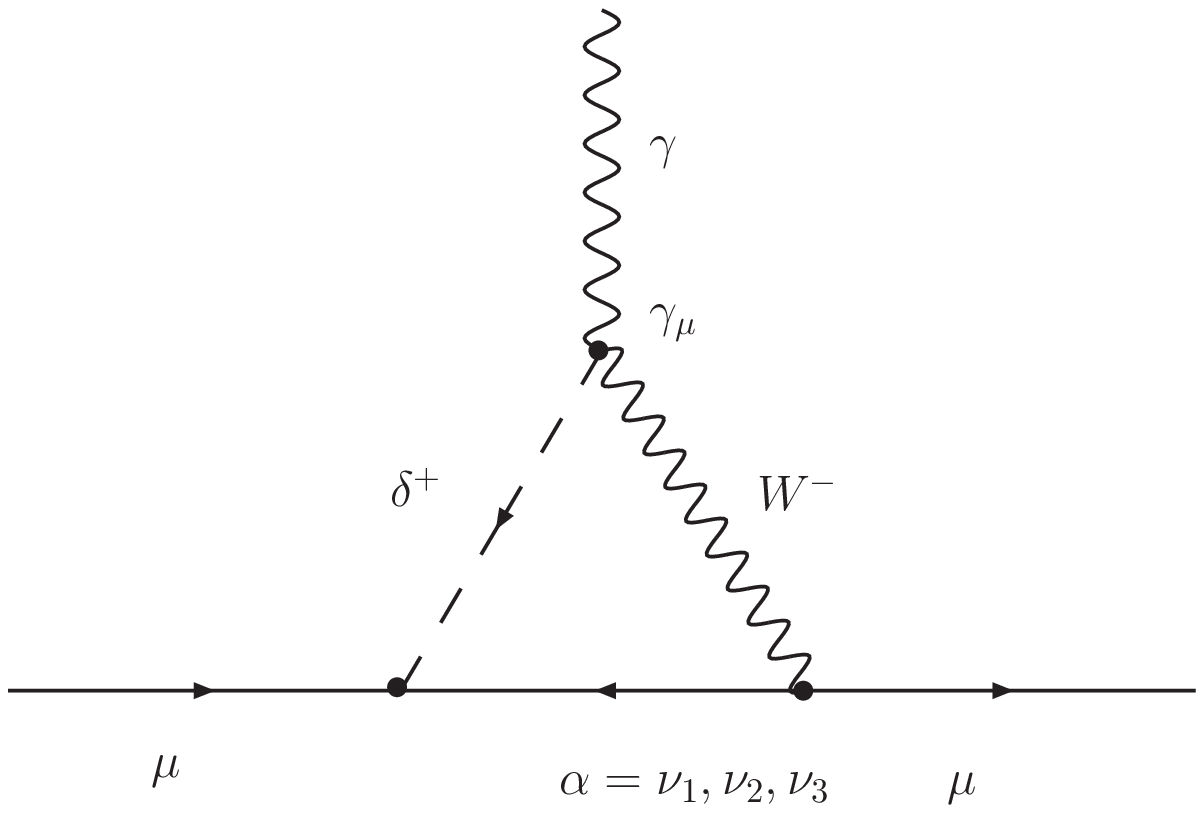,height=4cm,width=5.1cm,angle=0}}
\subfigure[]{\epsfig{file=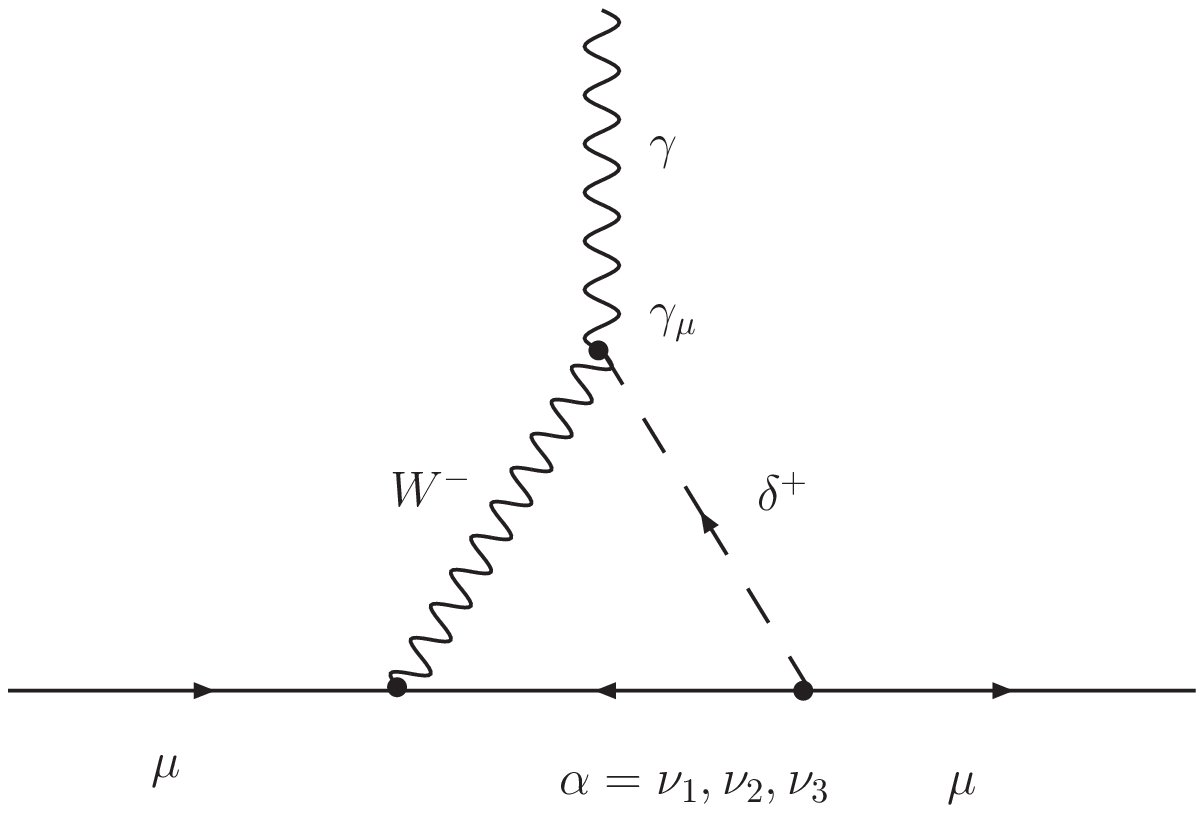,height=4cm,width=5.1cm,angle=0}}
\subfigure[]{\epsfig{file=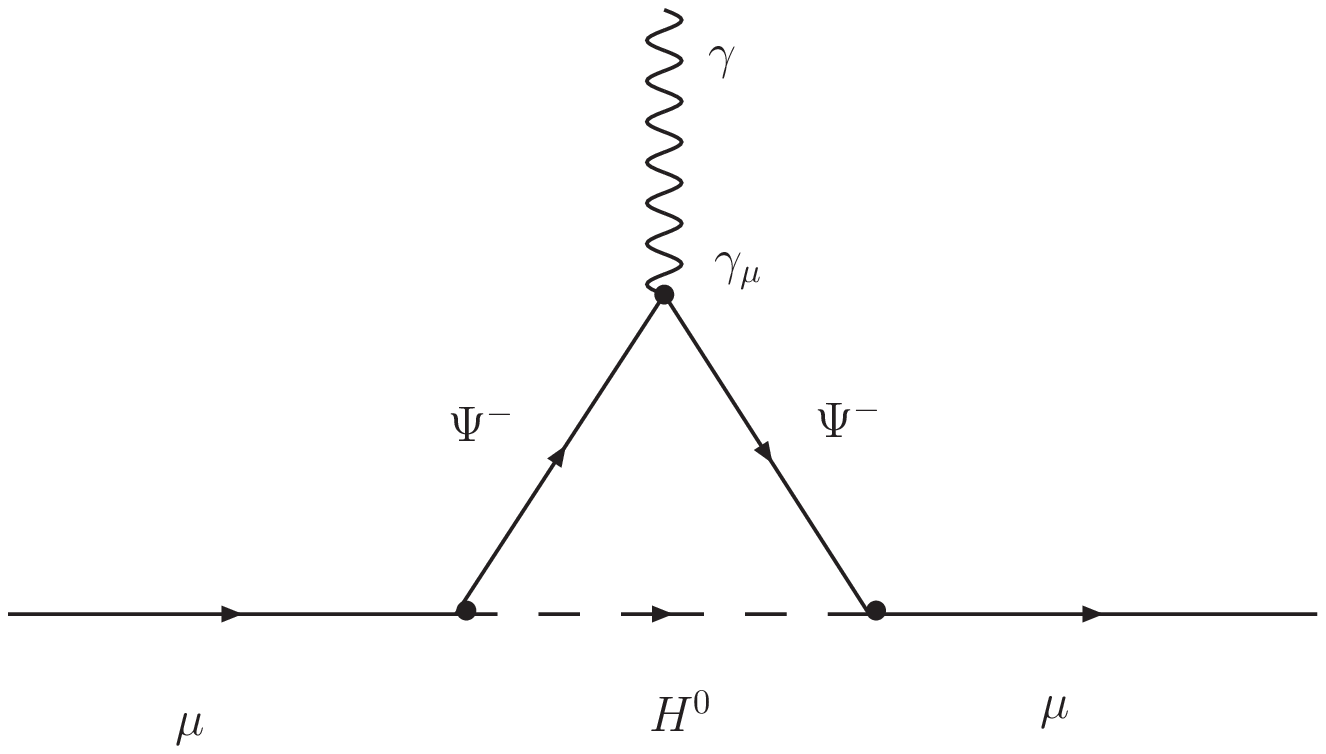,height=4cm,width=5.1cm,angle=0}}
\caption{One-loop Feynman diagrams contributing to $a_\mu^{}$.
Diagram (a) comes from the heavy neutrino and $W$ boson loop.
Diagrams (b) and (c) come from doubly charged Higgs and charged
lepton loops. Diagrams (d), (e), (f) come from singly charged Higgs
and neutrino ($W$ boson) loops. Diagram (g) comes from the triplet
fermion and SM-like Higgs boson loop.}
\end{figure}

\begin{figure}[t]
\epsfig{file=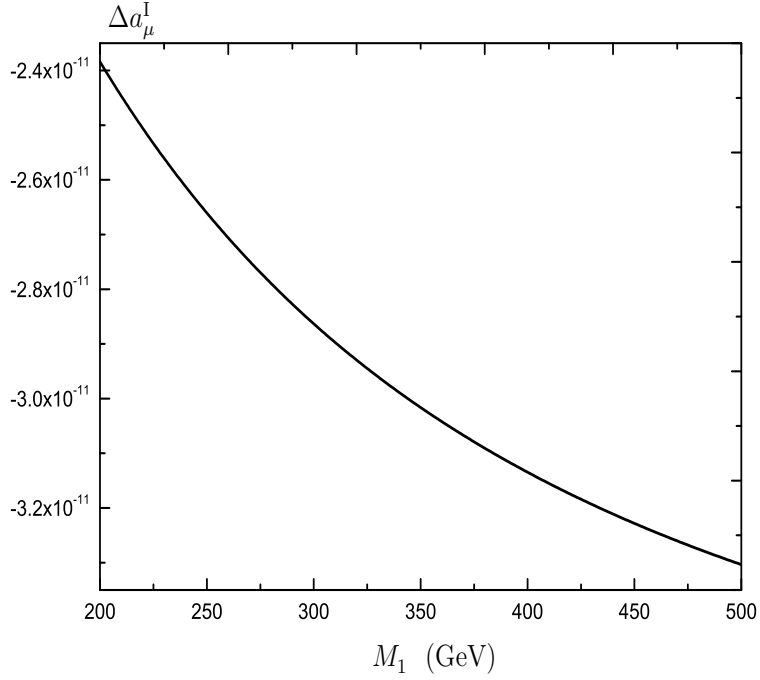,height=9cm,width=10cm,angle=0} \vspace{0cm}
\caption{$\Delta{a}_\mu^{\rm I}$ as a function of $M_{\rm 1}^{}$,
with $RR^\dagger\sim0.01$ and $200 {\rm GeV}\leq M_{\rm 1}^{} \leq ~
500 {\rm GeV}$.  }
\end{figure}

\begin{figure}[t]
\epsfig{file=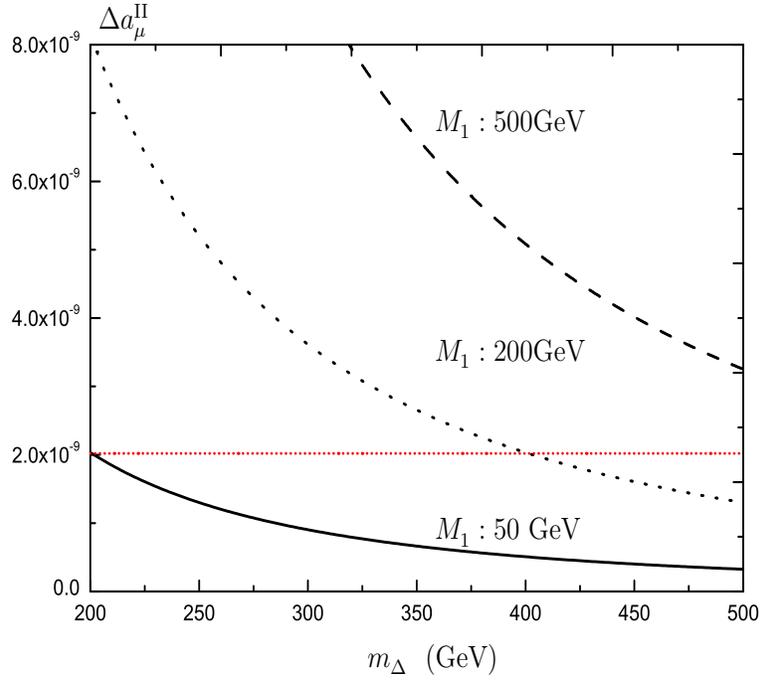,height=9cm,width=10cm,angle=0}
\caption{$\Delta{a}_\mu^{\rm II}$ as a function of $m_\Delta^{}$.
The solid, dotted and dashed lines correspond to $M_{\rm 1}^{}=50,~
200,~ 500~ {\rm GeV}$, separately. The short dotted line corresponds
to $\Delta a_\mu^{}$.}
\end{figure}

\end{document}